# HEAVY FERMION SUPERCONDUCTIVITY

by

Robert H. Heffner

Materials Science and Technology Division, MS K764

Los Alamos National Laboratory

Los Alamos , N. M. 87545

and

Michael R. Norman

Materials Science Division

Argonne National Laboratory

Argonne, IL 60439

# Contents

## I. Introduction

The study of heavy fermion (HF) superconductivity began about 16 years ago with the discovery of superconductivity in $CeCu_2Si_2$ [1]. At that time the major mystery was how superconductivity could be supported in a system with strong local moments on the Ce ions. Since that time the field of HF superconductivity has expanded to include five (possibly six) new uranium-based superconductors at ambient pressure and one new rare-earth-based superconductor at elevated pressures. Collectively, these materials display a rich variety of unexpected properties, which have enhanced our knowledge, not only of superconductivity, but also of the general behavior of highly correlated electrons in metals. It is interesting to note that the original question posed above for the discovery of $CeCu_2Si_2$, namely the coexistence of magnetism and superconductivity in HF materials, remains a central challenge for the understanding of these materials today. This paper addresses the question of how HF superconductors are different from conventional superconductors like Al and Pb and how these differences have enhanced our overall understanding of condensed matter physics. The topic is addressed within the context of four general areas: (1) normal-state properties, (2) the superconducting order parameter, (3) the interplay of magnetism and superconductivity, and (4) the superconducting pairing mechanism. The scope and depth of this article is intended to provide an easy, but stimulating experience for readers familiar with the basics of superconductivity. More comprehensive reviews of HFs has been given by Grewe and Steglich, Ott and Fisk, and Hess, Riseborough and Smith[1]. Normal state properties were reviewed in an earlier article in this series by Lawrence and Mills [2].

## II. Normal-State Properties

The starting point for understanding superconductivity in a material must be a study of the normal ground state from which the superconducting state emerges. Here we emphasize those normal-state properties (given in Table I) which are most useful in establishing a picture of the superconducting state.

The high-temperature ground state in HFs is reasonably well understood on the basis of extensions of single-ion Kondo theory [2]. How the variety of low temperature ground states (paramagnetic, magnetic, semiconducting and superconducting) evolve from the high-temperature normal state is only partially understood, however, although the following considerations are generally agreed upon. There is a crossover from local-moment behavior at high temperatures to a reduced-moment regime at low temperatures, where the f-electron moments are reduced to a fraction of their high temperature values. This compensation of the high-temperature f-moments occurs through an antiferromagnetic exchange interaction which produces a virtual bound state between the f-moments and the conduction electrons. In dilute metals this is the well-known single-ion Kondo interaction. Although this crossover from full- to reduced-moment behavior does not involve a phase transition, it can still be characterized by a coherence temperature $T_{coh} \approx$ 1-100K. Whether or not $T_{coh}$ is the only relevant energy scale at low temperatures in HFs is a matter of current investigation [3].

The strong exchange coupling between the conduction electrons and the f-electrons also leads to a large resistivity in HF materials, typically several hundred $\mu\Omega$-cm. at room temperature. The resistivity increases as the temperature is decreased below room temperature, reaching a maximum at $T_{max}$ and then dropping drastically below $T_{max}$, indicating the loss of inelastic scattering and the onset of the HF state. The normal-state resistivity at temperatures just above the superconducting transition temperature $T_c$ is typically only a few $\mu\Omega$-cm. At $T = T_{coh} << T_{max}$, if no magnetic ordering occurs, the system enters a Fermi liquid-like state with strongly renormalized quasiparticles of large effective mass m* $\approx$ 100 $m_e$. This state is characterized by a large linear coefficient of specific heat $\gamma$, indicating a high density of states at the Fermi surface, and a resistivity which varies as $T^2$.

The dc susceptibility in HFs is dominated by the f-electrons and also exhibits a maximum in temperature, below which it reaches a constant value if the f-moments are fully compensated. The zero-temperature values of $\chi$ and $\gamma$ for the family of HF materials appear to scale linearly with one another. It is interesting that among the HF materials the superconductors come closest to approaching the free-electron value (=1) for the Wilson scaling ratio R given by

$R = [\chi(0) \pi^2 k_B^2] / [\gamma(0) J(J+1) g_J^2 \mu_B^2]$.

Magnetic ordering in HF materials has almost always been found to be antiferromagnetic and occurs with moments reduced considerably over the free ion value (Table I), as expected from the screening by the conduction electrons. There appear to be four categories of magnetism. The first class ($U_2Zn_{17}$, $UCu_5$, $UPd_2Al_3$, $UNi_2Al_3$, $NpSn_3$) has moments ranging from 0.2 to 1.5 $\mu_B$ and appears to be composed of typical magnets with a standard dipolar order parameter. The second class ($URu_2Si_2$, $UPt_3$, $UPd_3$) has small moments (0.01 to 0.03$\mu_B$). This class is poorly understood at this time because, despite the possession of small moments, these materials still have Neel temperatures comparable to those in the first class. The most popular current model to explain these compounds is that the primary order parameter is quadrupolar (or a more exotic variant), with a weak induced dipole moment. The problem with this is that, so far, neutron scattering experiments have only detected the dipolar part. The third class of HF magnets contains those materials with extremely weak magnetism (0.001$\mu_B$) which is connected with a second superconducting phase transition, as seen in $UPt_3$ and Th-doped $UBe_{13}$. The most likely scenario to explain these small moments is a time-reversal-breaking superconducting order parameter, but other possibilities exist as discussed below. Finally, possible "spin glass," or random magnetic order, has been observed in $CeCu_2Si_2$ and $CeAl_3$, constituting a fourth class of HF magnetic materials.

Whether HFs exhibit an antiferromagnetic or paramagnetic ground state depends upon a competition between the local on-site exchange interaction, which compensates the local f-moment, and the non-local long-range f-f interaction, which gives rise to magnetic order through the RKKY interaction. The intra- and inter-site interactions are related through a common conduction-electron local-moment exchange coupling. Pethick and Pines [4] have shown that the competition between these two interactions may give rise to the coherence energy scale $T_{coh}$. In general, if $T_{coh}$ is very small one has local-moment behavior; by contrast, a large $T_{coh}$ favors a paramagnetic ground state. Even in the paramagnetic state, however, neutron scattering experiments [5] have demonstrated that HFs exhibit strong antiferromagnetic (AFM) spin correlations between f-moments.

It is evident that the normal-state properties of HF materials are very different from conventional intermetallic compounds. The principal differences discussed here probably result from a delicate balance between competing interactions, giving rise to a highly-correlated ground state which at low temperatures exhibits a small resistivity, a large effective mass, strong AFM spin correlations and highly-reduced ordered moments. As discussed below, these properties have profound implications for the superconducting state. The reason that only a relatively small family of materials exhibit these unusual characteristics results from a combination of the strong exchange interaction and an optimum f-ligand and/or f-f atom spacing which tunes the competition between the inter- and intra-site couplings.

### III. Superconducting Order Parameter

A superconductor like Al is very well described by the well-known BCS model of superconductivity, originally constructed for a weak pairing interaction between electrons in a relative zero angular momentum, spin-singlet state. This implies a spherically symmetric order parameter $\Delta(\mathbf{k})$, which is the pair amplitude for electrons of momentum $\mathbf{k}$. In general, because of crystalline inversion symmetry, $\Delta(\mathbf{k})$ can be written separately for an even or odd parity pairing state [6]:

$$\Delta(\mathbf{k}) = \psi(\mathbf{k}) \, i \sigma_y \tag{1a}$$

$$\Delta(\mathbf{k}) = (\sigma \cdot \mathbf{d}(\mathbf{k})) \, i \sigma_y, \tag{1b}$$

respectively. Here $\psi(\mathbf{k})$ and $\mathbf{d}(\mathbf{k})$ are even scalar and odd vector functions of the momentum $\mathbf{k}$ and $\sigma$ is the Pauli spin matrix. (In the presence of strong spin-orbit coupling spin is replaced by pseudo-spin, as discussed in Section VII below.) The BCS state in Al is described by Equation (1a). The symmetry operations of the Hamiltonian include translation symmetry, inversion symmetry (parity), point group symmetry, time-reversal symmetry and gauge symmetry. When a superconducting transition occurs only gauge symmetry, which sets the overall phase of the order parameter, is necessarily broken. It is customary to refer to an "unconventional" superconducting state as one possessing broken point group symmetry, having odd parity or violating time-reversal symmetry. The latter occurs when the superconducting state possesses either orbital or spin moments. In addition, if the Cooper pairs have finite center-of-mass momentum, an additional translational symmetry will be broken. Finally, if one has a triplet superconductor (Eqn. 1b) and no spin-orbit coupling, then another symmetry connected with the rotation group of the spins can be broken. This leads to a large number of possibilities. These points are discussed in greater detail in Section VI below.

When Δ(**k**) vanishes for certain **k**-vectors one measures specific power-law temperature dependencies for all experiments involving the thermal excitation of quasiparticles near the nodes of Δ(**k**) at $T \ll T_c$. This is illustrated in Table II for polar and axial gap structures on a spherical Fermi surface. A gap which is non-vanishing for all **k** will yield an exponential temperature dependence. As discussed below, the observation of power-law temperature dependencies below $T_c$ is not conclusive evidence for an unconventional gap structure, however, because gapless superconductors can also exhibit power-law behaviors. More stringent tests for unconventional superconductivity involve the detection of more than one superconducting phase, an inherently magnetic superconducting state (violating time reversal symmetry) or anisotropy in the critical fields, transverse ultrasound, thermal conductivity or penetration depth, for example [7].

## IV. Superconducting Properties of Heavy Fermions

*A. General Thermodynamic Properties*

The six known HF materials which are superconducting at ambient pressure are shown in Table III, together with some of their relevant properties. All are type II superconductors. Also included for comparison are two elemental superconductors, Al and Nb, and an A-15 compound $V_3Si$. A seventh HF superconductor, $CeCu_2Ge_2$, has been reported to be superconducting above 70 kbar pressure by Jaccard et al.[8]. In addition, $U_2PtC_2$ is probably a HF superconductor [9].

Several trends are apparent from Table III. All of the HF superconductors have small transition temperatures ($\leq$ 2K). In addition, the HFs have magnetic field penetration depths λ in excess of several thousand Å and coherence lengths ξ of the order of 100-200 Å. This is in contrast to the conventional elemental superconductors which have significantly lower values of λ and higher values of ξ. These differences are a direct consequence of the very large effective masses for the HFs (Table I), as can be seen by examining the expressions for λ and ξ (taken from conventional theories of superconductivity): $\lambda \propto (m^*/n_s)^{1/2}$ and $\xi \propto n^{1/3}/(m^*T_c)$, where $n$ is the conduction electron density and $n_s$ is the superconducting pair density. This leads to anomalously small values of $H_{c1} \propto 1/\lambda^2$ and large values of $H_{c2} \propto 1/\xi^2$ when compared to the thermodynamic critical fields $H_c$ of the elemental type I superconductors. Note that if one calculates $H_c = (H_{c1}H_{c2})^{1/2}$ for the HF superconductors, however, the values are *not* anomalously large compared to the elemental superconductors. Also, the $H_{c2}$ values for the HFs are rather *small* compared to the A-15 superconductors like $V_3Si$ ($H_{c2}$ = 19T) or to the copper-oxide superconductors, where $H_{c2}$ > 100T. The small coherence length for $V_3Si$ derives from a moderately large effective mass and a large $T_c$. (A discussion of the unusual properties of the A-15 materials is beyond the scope of this article. The example of $V_3Si$ is included here only for comparison.) Finally, the HF superconductors exhibit rather conventional specific heat jumps $\Delta C$ at $T_c$ compared to Al, where the BCS value for $\beta = \Delta C/\gamma(T_c)T_c$ is 1.43. (An exception is $UBe_{13}$, where $\beta$ can be up to 2.6 for single crystals.) The parameter $\beta$ is a measure of the pair coupling strength. Thus the uniqueness of HF superconductors cannot be established from the magnitudes of their thermodynamic parameters alone.

*B. Experimental Evidence for Unconventional Superconductivity*

The earliest evidence for an unconventional gap structure in HF superconductors comes from the measurement of power-law temperature dependencies for the resistivity, specific heat, sound attenuation and NMR spin-lattice-relaxation rate below $T_c$ [1]. In no HF superconductors did these early measurements yield the exponential behavior expected for a gap without nodes on the Fermi surface. The problem is that these experimental data, while suggesting an anisotropic gap structure, do not uniqely determine the gap symmetry. One reason for this is that specific power laws (Table III) are valid only far below $T_c$ where impurities can obscure the intrinsic behavior. Another is that certain measurements may be dominated by one feature of the gap structure, such as a line of nodes. This could explain why the NMR spin-lattice-relaxation rates show a $T^3$ behavior over a wide temperature range below $T_c$ for <u>all</u> of the HF superconductors. If point nodes were also to exist (as postulated for $UPt_3$), for example, they may not affect the NMR rates because of a smaller phase space and because the experiments are generally performed on powder samples where directional information can be lost.

Another indication of unconventional superconductivity in HFs occurs when dilute concentrations of substitutional impurities are introduced. In conventional s-wave superconductors, magnetic impurities with stable local moments (such as Gd) cause exchange scattering which breaks the Cooper pairing and yields a universal curve for $T_c$ vs. the impurity concentration or for the specific heat jump $\Delta C/\Delta C_0$ vs $T_c/T_{c0}$. This effect is well described by the theory of Abrikosov and Gor'kov [10]. By contrast, dilute concentrations of non-magnetic impurities in s-wave superconductors have little effect on the superconducting parameters. This is not the case for HF superconductors, however. In these systems non-magnetic impurities (such as Y, La or Th) are nearly as effective at breaking pairs as are magnetic impurities like Gd. This has been interpreted as evidence for non s-wave pairing, because non-magnetic impurity scattering suppresses an anisotropic Cooper pair wave function. When an impurity such as Gd is introduced into an anisotropic superconductor, magnetic impurity scattering produces the additional pair breaking described by Abrikosov and Gor'kov.

Although the power-law behaviors and sensitivity to non-magnetic impurities are indicators of an anisotropic pairing state in HFs, the first definitive evidence for unconventional superconductivity in undoped HF systems came only recently with the discovery of two superconducting transitions in the specific heat of $UPt_3$ by Fisher et al.[11], as shown in Fig. 1. The earliest specific heat measurements in $UPt_3$ missed this small splitting ($\Delta T \approx 0.05K$), apparently because of sample quality and less refined annealing procedures. This splitting of the specific heat jump in $UPt_3$ was first discovered in zero applied field. Subsequent studies, most notably changes in ultrasonic velocity in applied fields of several Tesla or more by Bruls et al. and Adenwalla et al. [12], have revealed a rich field-temperature phase diagram for $UPt_3$, displaying at least three different superconducting phases and a tetracritical point at ambient pressures. Surprisingly, it was found that the phase diagram is not very sensitive to field orientation. In particular, the same tetracritical point exists for fields along the c-axis, where no kink is observed in the upper critical field. Recently, Boukhny et al. [13] have extended the sound velocity measurements to pressures of several kbars, showing how these phases evolve under pressure. The phase diagram for $UPt_3$ is shown in Figs. 2 and 3, and constitutes prima facia evidence for an unconventional superconducting state. The low-field transition from the A to the B phase in $UPt_3$ at atmospheric pressure has recently been shown with muon-spin-relaxation ($\mu$SR) measurements by Luke et al. [14] to be accompanied by the onset of weak magnetic ($\mu \approx 0.001\ \mu_B$) correlations. In addition, Keller et al. [15] have found a six-fold anisotropy in the basal-plane magnetoresistance of $UPt_3$, and Broholm et al. [16] used μSR to measure a different temperature dependence for the magnetic field penetration depth λ, depending upon the orientation of the applied field. These penetration depth measurements are consistent with previous transverse ultrasound studies by Shivaram et al. [17], and indicate line nodes perpendicular to the c-axis. Finally, Kleiman et al. [18] have observed the hexagonal vortex lattice in UPt$_3$ using neutron diffraction. The implications of many of these experiments for the symmetry of $\Delta(\mathbf{k})$ will be discussed in Section VI.

Pure UBe$_{13}$ apparently exhibits a single superconducting phase transition, although there have been reports of structure in $H_{c2}(T)$ below $T_c$ by Rauchschwalbe et al. [19], possibly indicating a second phase. No clear evidence for two phases has been seen in the specific heat, however, which sas shown by Ott and collaborators to have a low-temperature behavior proportional to $T^3$ [20]. Einzel et al. [21] have found that the magnetic field penetration depth varies at $T^2$ at low temperatures. These two results taken alone imply an axial state with point nodes, as seen from Table II. MacLaughlin's NMR relaxation rates are proportional to $T^3$ at low temperatures, however, which taken by itself implies a polar state with a line of nodes [22]. Even though the NMR data were taken in 1.5T field (and therefore may not be comparable to the zero-field specific heat data), this comparison illustrates the problem of using power-law temperature dependencies to determine the gap symmetries. The fact that NMR has yielded a $T^3$ behavior in all HF superconductors, as discussed above, further compounds the problem.

When either magnetic (Gd) or non-magnetic (La) impurity atoms are substituted for U in UBe$_{13}$, $T_c$ is depressed as expected for a superconductor with an anisotropic order parameter. However, when doped with Th, Ott et al. found that the $T_c$ for $U_{1-x}Th_xBe_{13}$ becomes a non-monotonic function of Th concentration x and for $0.019 \leq x \leq 0.043$ two specific heat transitions occur [23], as seen in Fig. 4. Early ultrasonic attenuation measurements by Batlogg et al. gave an indication for antiferromagnetism at the lower transition [24]. The existence of magnetic correlations was later confirmed in zero-field μSR experiments by Heffner et al. [25], revealing a mean-field like magnetic transition below $T_{c2}$, with an approximate moment $\leq 0.001$ μ$_B$. No such signature was found at $T_c$, the onset of superconductivity, as seen in Fig. 5. In addition, $H_{c1}(T)$ was found to be proportional to $(1-\alpha(T/T_c)^2)$ above and below $T_{c2}$, but with a larger slope α below $T_{c2}$ than above it [25], implying an increase in the superfluid density as expected for a second superconducting transition. Finally, Jin et al. [26] showed that the addition of Th to $UBe_{13}$ can produce a relatively large zero-temperature specific heat coefficient $\gamma_r(0)$ of order 0.5 - 1.0 $J/mol\text{-}K^2$, and can change the temperature dependence of the low-temperature specific heat.

An important question arises: Is the lower transition in (U, Th)$Be_{13}$ associated with a transition to a different superconducting state (possibly violating time-reversal symmetry) or is it simply associated with a small-moment AFM transition? Part of the answer lies in the large specific-heat jump at $T_{c2}$. Simply put, a small local-moment AFM transition is unlikely to produce such a large jump in specific heat because most of the entropy associated with the exchange-coupled conduction-electron/local-moment system would already have been removed in reducing the value of the moments to $0.001\mu_B$. If the magnetic order occurs through a SDW transition (a Fermi surface instability, rather than local moment formation), the large specific heat jump would again be unlikely because the Fermi surface would already have been largely consumed by the superconducting transition at $T_c$. For a SDW transition, in other words, a large jump would require an exceptional enhancement of the density of states near the zeros of the superconducting energy gap. We thus conclude that the large change in specific heat at $T_{c2}$ is due primarily to a superconducting, not a magnetic, phase transition. Nevertheless, it is still true that magnetic correlations are also observed to set in at $T_{c2}$. The possible nature of the magnetism and the implications for the nature of the superconductivity in $UBe_{13}$ are discussed below in Sections V and VI.

Further confirmation for multiple superconducting states in (U,Th)$Be_{13}$ comes from Lambert's early measurements of the pressure dependence of $T_c$ for various Th concentrations [27]. The functional $dT_c/dP$ changes from about -16 mK/kbar for $0.0 \leq x \leq 0.017$ to about -50 mK/kbar for $0.019 \leq x \leq 0.026$, indicating different superconducting phases for different range of Th concentrations. It was also found that higher pressures for $0.019 \leq x \leq 0.026$ seemed to restore the phase seen for $x \leq 0.017$.

The combined specific heat, critical field and µSR studies led to the phase diagram [25] shown in Fig. 6 for (U, Th)$Be_{13}$. The vertical dotted lines were not actually been observed, however, but only postulated from measurements for various Th concentrations, differing by several tenths of a per cent Th. As discussed below, Th tends to act a as "negative pressure" when substituted in $UBe_{13}$. Recently, Zieve *et al.* [28] measured the specific heat as a function of pressure for x = 0.022 and confirmed the existence of the nearly vertical phase boundary near x = 0.019 in the T vs. x phase diagram. Here the pressure was tuned to achieve an equivalent resolution in Th concentration of 0.0002. They found that the relative specific heat jump $\Delta C/C$ decreases by about 10% in going across the A $\rightarrow$ C phase boundary at constant temperature (Figure 6). This is to be contrasted with an increase in $\Delta C/C$ of about 50% in decreasing the temperature at constant pressure and traversing the B $\rightarrow$ C phase boundary.

Thus, both $UPt_3$ and $(U,Th)Be_{13}$ exhibit multiple superconducting transitions, providing unambiguous evidence for unconventional superconductivity in these HF systems. The origins of the magnetism below $T_c$ and the symmetry of the superconducting states in these systems are discussed in Sections V and VI below.

*C. Interplay of Magnetism and Superconductivity*

It is evident from the above discussion of the multiple superconducting phases of $UPt_3$ and $(U,Th)Be_{13}$ that magnetism and superconductivity seem to be intimately related in HF systems. Indeed, where magnetic or superconducting phase transitions are found in HFs, it is clear from the entropy balance above and below the transition that the heavy quasiparticles themselves (ie., the hybridized f-electron/conduction-electron states) form the ordered magnetic or superconducting state. This fact is in contradistinction to a previous class of "magnetic superconductors," the moly-sulfides and rare-earth rhodium borides, where the magnetism is carried by the local rare-earth moments but the superconductivity occurs through separate conduction-electron bands which interact only weakly with the local moments. It is also worth noting at this point that there is growing evidence, discussed below, that magnetic spin fluctuations (paramagnons) may provide the dominant pairing mechanism in HF superconductors.

This interplay between magnetism and superconductivity in HF materials has shown considerable variety when viewed form the point of view of the competition, coexistence and/or coupling of the magnetic and superconducting order parameters [29]. As seen from Table I magnetism can occur both above and below the superconducting transition temperature.

*C1. Competition*

In $CeCu_2Si_2$, the only rare-earth-based HF superconductor at ambient pressure, spontaneous magnetism occurs slightly above $T_c \cong 0.7K$ and exists along with superconductivity in zero-applied field. In non-zero fields the magnetism persists above $H_{c2}$ for temperatures below 0.7K, as seen in the phase diagrams formulated by Nakamura, Wolf and collaborators [30]. Only one superconducting phase has been established. Experiments using μSR, specific heat and microstructural analysis by Luke et al. and Feyerherm et al. provide insight into the interplay of magnetism and superconductivity in $CeCu_2Si_2$ [31]. The zero-field μSR relaxation function in $CeCu_{2.05}Si_2$ exhibits a two-component structure below 1.3K, where magnetism sets in, one component of which can be attributed to paramagnetic domains and the other to magnetic domains. The magnetic component exhibits a Gaussian form, characteristic of either a spin-glass-like state or an incommensurate SDW. This segregation occurs in an otherwise homogeneous sample. It is therefore possible to track the temperature dependence of the volume fractions of the magnetic and non-magnetic portions of the sample, as shown in Figure 7. The magnetic fraction increases below 1.3K, reaching a maximum of about 3/4 at $T_c$. Below $T_c$ the magnetic fraction decreases to about 1/3 at $T \leq 0.05K$, while the superconducting (paramagnetic) volume fraction increases to about 2/3. Corroborating evidence for this effect is found in Nakamura's NMR measurements [32] on $CeCu_{2.02}Si_2$, where the NMR intensity decreases strongly below about 0.9K due to the onset of magnetic correlations which produce a broad distribution of hyperfine fields. This decrease in intensity stops at $T_c$, where the μSR measurements show that the magnetic volume fraction ceases to increase. Also, the height of the specific heat jump at $T_c$ correlates well with the fraction of the sample which is superconducting, as obtained from the μSR measurements. This effect has been studied in several different samples with varying Cu stoichiometries, where Cu excess tends to favor superconductivity. In each case the magnetic relaxation rate is unaffected by the onset of superconductivity and is relatively independent of the sample, indicating a common origin from one sample to the next.

Thus in $CeCu_2Si_2$ there appears to be a competition between magnetism and superconductivity, each existing in its own domains. This effect is unique among HF superconductors (as discussed below) and may be related to the fact that the existence of magnetism and/or superconductivity in $CeCu_2Si_2$ is very sensitive to subtle changes in stoichiometry, which may be related to changes in unit-cell volume or internal strain. Extensive experiments inducing volume changes have been carried out by La doping and Cu deficit (where $\Delta V/V > 0$) or Cu excess and hydrostatic pressure (where $\Delta V/V < 0$) [1]. Samples with large unit-cell volume tend to be magnetic and non-superconducting, reflecting a reduction of $T_{coh}$. When the volume is reduced, $T_{coh}$ is raised and superconductivity is found.

*C2. Coexistence*

The next three HF superconductors in Table III, $URu_2Si_2$, $UPd_2Al_3$ and $UNi_2Al_3$, all exhibit coexistence of superconductivity with magnetic order which sets in above $T_c$. $URu_2Si_2$ [33] forms a body-centered tetragonal structure and displays small-moment AFM ordering ($\mu = 0.03\mu_B$) at $T_N = 17K$ with its moments polarized along the tetragonal c-axis. Neutron scattering (Mason et al.[34]), x-ray scattering (Isaacs et al. [35]) and $\mu$SR (Knetsch et al. [36]) all provide strong evidence that this AFM order persists below the superconducting transition. A similar coexistence is found by Geibel and Feyerherm and collaborators in the hexagonal compound $UPd_2Al_3$, except that the ordered moment in this system is larger, $\mu = 0.85\mu_B$, and $T_N = 15K$ [37]. Here the moments are aligned ferromagnetically in the basal plane, alternating directions from one basal plane to the next. $UNi_2Al_3$ has the same lattice structure as $UPd_2Al_3$, but displays incommensurate AFM, with an ordering wave vector given by $(1/2 \pm 0.11, 0, 1/2)$ (Geibel, Schroder et al. [38]). As in $UPd_2Al_3$ the moments lie along the a-axis in the basal plane, alternating antiferromagnetically along the c-axis. However, the moment is only $\mu \approx 0.24\mu_B$, but again $\mu$SR measurements by Amato and collaborators [39] provide strong evidence for persistence of the magnetic ordering in the superconducting state.

It is important to note that the existing data show that the magnetism and superconductivity in $URu_2Si_2$, $UPd_2Al_3$ and $UNi_2Al_3$ coexist on a microscopic scale. For example, as displayed in Fig. 8 for $UPd_2Al_3$, the $\mu$SR relaxation rate increases below $T_N$ and then again below $T_c$, due first to the onset of AFM and then to the field inhomogeneity produced by the superconducting flux lattice, respectively. Because the data display only a single relaxation component which responds to both the superconducting and magnetic behavior, microscopic coexistence is implied. Furthermore, although there is microscopic coexistence of these two ground states, the data show that they do not appear to strongly interact with one another, despite the fact that the f-electrons are involved in each state. This is different from the case of $CeCu_2Si_2$, described above, and from $UPt_3$ and $(U, Th)Be_{13}$, described next.

*C3. Coupling*

$UPt_3$ undergoes a small moment ($\mu = 0.02\mu_B$) magnetic phase transition with the spins ordered antiferromagnetically in the basal plane below $T_N = 5K$ [40]. Unlike the three U-based superconductors discussed above, however, the magnetic order in $UPt_3$ is affected by the onset of superconductivity. This is seen in the neutron scattering experiments of Aeppli et al. (and most recently Isaacs et al.) as a decrease in the magnetic Bragg intensity below $T_c$ in zero applied field [41]. The interpretation of this phenomenon is in terms of a coupling between the superconducting and magnetic order parameters which reduces the sublattice magnetization with the onset of superconductivity. Further evidence for this has come from a beautiful comparison of the magnetic Bragg intensity and the splitting in the superconducting specific heat anomaly as a function of pressure carried out by Hayden et al. [42]. It is found that the upper and lower transition temperatures in $UPt_3$ merge at a pressure of about 3.5 kbars, about the same pressure where the magnetic moment observed at 5K also vanishes (Fig. 9). Above 4 kbar there is no moment and no splitting. Moreover, the $T_c$ splitting is proportional to the square of the magnetic moment. A simple interpretation is that the magnetic moment causes the splitting by breaking the hexagonal symmetry in the basal plane. This coupling would set limitations on the symmetry of $\Delta(\mathbf{k})$, as discussed below.

This picture of a magnetic symmetry-breaking field in $UPt_3$ has been challenged, however, by noting that the coherence length for the small moment ordering at 5K is only a few hundred Å, of the order of the superconducting coherence length itself. Whether this lack of long-range coherence in the magnetic order parameter is severe enough to disqualify the 5K magnetism from playing the role of a symmetry-breaking field is unclear. For example, at a microscopic level the effect of the symmetry-breaking field would be on the pairing interaction, whose range is much smaller than the coherence length. In fact, it may be that it is actually the low-frequency spin fluctuations which act as the symmetry breaking field (as opposed to the small ordered moment). Such a scenario would then be analogous to the strong coupling stabilization of the A phase in $^3He$, and would support a magnetic origin for the broken symmetry. Nevertheless, it has been suggested that structural disorder may be the source of both the magnetic order and the symmetry-breaking field which splits the superconducting transition. Recent TEM measurements [43] by Midgely et al. on thin slices of $UPt_3$ have revealed an incommensurate lattice distortion with a modulation of about 50Å and a long-range coherence of the order of $10^4$Å. Because this structure would locally lower the hexagonal symmetry (to trigonal) it could, in principle, couple to the superconducting order parameter in a manner analogous to the magnetic coupling discussed above. Indeed, experiments have demonstrated a degradation of the long-range coherence of the microstructure and the loss of the sharp two-peak structure in the specific heat in unannealed samples, indicating a relationship between the microstructure and the superconductivity. Whether this microstructure is present in the bulk (x-rays have not seen it) and whether it can be destroyed by only 4 kbars of pressure (it has a high formation temperature) are open questions, however.

## V. Magnetism in HF Superconductors

*A. Magnetism Originating Below $T_c$*

The coexistence of magnetic correlations and superconductivity in $(U,Th)Be_{13}$ was mentioned above. The normal state properties of $(U,Th)Be_{13}$ may give a clue to the origin of this weak magnetic signature in the superconducting state. In pure $UBe_{13}$ the resistivity $\rho$ passes through a maximum around $T_{max}$ = 2.2K before decreasing strongly, the latter signaling the onset of the coherent scattering regime. (Of course, below $T_c$ = 0.9K $\rho$ goes to zero.) Borges et al. showed that the resistivity maximum at $T_{max}$ decreases in temperature as the Th content is increased, such that $T_{max}$ intersects the superconducting $T_c$ vs x phase diagram at x ≈ 0.019 [44]. This is just where the second phase transition at $T_{c2}$ occurs, where the upper critical temperature $T_{c1}$ begins to increase with x, and where magnetic correlations are seen by $\mu$SR below $T_{c2}$. Also $T_{max}$ falls below the superconducting critical temperature for x > 0.019 at ambient pressures. This latter fact has been established by measuring the pressure dependence of the resistivity $\rho(T,x)$. These pressure measurements [44] also show that $\rho$ vs T scales into a universal curve of $\rho/\rho_{max}$ vs. $T/T_{max}$, indicating the presence of a new low-temperature energy scale given approximately by $T_{max}$. Here $\rho_{max}$ is the value of the resistivity at $T_{max}$. The magnetic field dependence of $\rho_{max}$ was measured by Knetsch et al. [45] and suggests a magnetic origin for this feature, possibly caused by the scattering of conduction electrons from the correlated fluctuations of the uranium spins. This idea is qualitatively consistent with theoretical models of the normal state of heavy fermion systems, which show that a new energy scale can emerge from the intersite exchange coupling between Kondo ions [4]. The point here, however, is that the weak magnetism observed below $T_{c2}$ could be due to the freezing out of the spin fluctuations as $T_{max}$ falls below the superconducting transition temperature. If so, it is noteworthy that the line of phase transitions at $T_{c2}$ begins and ends on the line of superconducting phase transitions at $T_{c1}$ (for $x \approx 0.019$ and $x \approx 0.043$, respectively), thus indicating a coupling between magnetic and superconducting order parameters in this system. Also, the magnetism would have to be accompanied by a change in the superconducting state at $T_{c2}$, as argued above in Section IV B.

The preceding discussion assumed that the observed magnetic correlations in $(U,Th)Be_{13}$ are due to AFM of some sort. Whether this is true, or whether the magnetism arises from a magnetic superconducting state, is still being investigated. (This is also true of the magnetic correlations observed in the low-field A to B transition in $UPt_3$, mentioned above.) This second option will be discussed more below, in the section on the symmetry of the HF superconducting state.

There is a third option which also deserves investigation as a possible source of the small-moment magnetism found below $T_c$ in at least some of the HF superconductors. It has been shown theoretically by many authors [46] that the presence of non-magnetic scattering centers can lead to a resonance state below the gap energy whose density of states increases as the temperature is decreased below $T_c$. For an isotropic gap Balatsky *et al.* have shown [47] that the resonance wave function falls off exponentially as $\psi \approx \exp(-r/\xi)$, where $\xi$ is the superconducting coherence length. For an unconventional superconductor with line nodes, however, $\psi$ falls off more slowly along the direction of the nodes ($\psi \approx 1/r$), and exponentially elsewhere. A characteristic signature of this enhanced density of states is a large residual linear specific heat coefficient $\gamma_r(0)$ at T = 0. Interestingly, both $UPt_3$ [11] and $(U,Th)Be_{13}$ [26] possess large $\gamma_r(0)$ values where μSR experiments see small-moment magnetism below $T_c$ ($\gamma_r(0) \cong 0.5 - 1.0$ J/mol-$K^2$ for Th-doped $UBe_{13}$). This correlation has not been completely established, however, because the existence of a large $\gamma_r(0)$ in $UPt_3$ is apparently sample-dependent, and a large $\gamma_r(0)$ is seen in $U_{1-x}Th_xBe_{13}$ for x = 0.06, where no enhanced μSR signature is observed [25,26]. (In the latter case, disorder could possibly weaken or destroy the quasistatic magnetism, however.) Nevertheless, further theoretical and experimental investigations of possible magnetic ordering arising from resonant impurity scattering are worthwhile. A recent case in point involves the study of $UBe_{13-y}B_y$, first investigated by Felder, Ott and coworkers [48]. Here the B substitutes for the Be atoms at the cubic sites in $UBe_{13}$ and drastically alters the behavior of the specific heat [48, 49]. For example, the normalized specific heat jump is quite large when y = 0.03 ($\Delta C/\gamma(T_c)T_c \cong 4.4$), indicating a change in the pairing strength. In addition, $\gamma_r(0) \cong 0$ and no evidence for small-moment magnetism is found for y = 0.03. For y = 0.11, however, $\gamma_r(0) \cong 1$ J/mol-K$^2$ and recent μSR experiments by Heffner et al. [50] have found a magnetic signature below $T_c$. Furthermore, these experiments and those of Beyermann et al. [49] have shown that boron-doped $UBe_{13}$ shows only one specific heat jump for all B concentrations measured to date (y ≤ 0.11), unlike $(U,Th)Be_{13}$. Thus this correlation between $\gamma_r(0)$ and small-moment magnetism below $T_c$ deserves further study.

*B. Magnetism Originating Above $T_c$*

If these magnetic signatures are caused by ordinary AFM, however, there is a natural way of understanding the prevalence of magnetism which coexists in all of the U-Based HF superconductors. The starting point is that the strongly hybridized f-electron quasiparticles are involved in both the magnetic and superconducting phases of these materials. Thus, as Kato and Machida showed sometime ago, the f-electron Fermi surface could be shared between superconductivity and AFM, such that spin-density-wave (SDW) nesting occurs where there are nodes in the superconducting order parameter [51]. This requires an unconventional (anisotropic) superconducting state, which is prevalent in HF systems. This sharing of the Fermi surface seems to be the case in $URu_2Si_2$, for example, where Palstra et al. showed that the size of the specific heat jumps at $T_N$ and $T_c$ indicate that neither phase transition consumes the full Fermi surface by itself [33]. It should be noted, however, that the small-moment magnetism in $URu_2Si_2$ and $UPt_3$ above $T_c$ is probably of a different origin than the extremely small-moment magnetism associated with the superconducting phase in $UPt_3$ and Th-doped $UBe_{13}$.

The notion that magnetism and superconductivity might share the Fermi surface, as mentioned above, assumes that the magnetic order arises from a Fermi surface instability, as in a SDW. This, of course, implies that the f-electrons producing the ordered moment are largely itinerant. This seems to be the case in the HF superconductors $UPt_3$, $URu_2Si_2$, and $UNi_2Al_3$, for example, but not in $UPd_2Al_3$ (even though band calculations [52] give a good description of the magnetic phase). In the latter system recent studies of the resistivity and specific heat at ambient and applied pressures up to about 11 kbars [Caspary *et al.* in Ref. 37] indicate that the AFM order observed at $T_N$ = 14K is due to localized f-electrons. The evidence for this comes from the observation of crystal-field excitations and from the fact that it is not necessary to invoke a gap in the electronic density of states to explain the resistivity data near $T_N$. (Such a gap is observed in $URu_2Si_2$ and $UNi_2Al_3$ and is characteristic of a Fermi surface instability in an itinerant electron system). Furthermore, the data imply that the f-electron density of states may be split into a local moment part and a less-localized part, the latter giving rise to moderately heavy electron behavior and a superconducting transition near 2K. Thus in $UPd_2Al_3$ the coexistence of magnetic order and superconductivity is of a somewhat different nature than in $UNi_2Al_3$, for example. This contrast emphasizes the need to study the distinguishing normal and superconducting properties of each HF superconductor before drawing general conclusions.

Interactions between superconductivity and magnetism can be described theoretically through terms in the free energy of the system which couple the vectors representing the superconducting and magnetic order parameters, while preserving the overall point group symmetry of the Fermi surface. These symmetry considerations are described in the next section.

### VI. Phenomenological Models for the Superconducting Order Parameter

A good review of this area has been given by Sigrist and Ueda [53]. Our purpose here is to review current developments which attempt to construct models for the order parameter consistent with the available experimental data. We first give a brief overview of how the order parameter is constructed. To begin with, we note that because of the assumption of strong spin-orbit coupling in HF systems, the spin and orbital degrees of freedom are no longer separately good quantum numbers. One can still construct eigenstates of a pseudo-spin operator, however, using the fact that Kramers theorem makes each **k**-state two-fold degenerate (corresponding to 'spin up' and 'spin down'). The operation which connects these two states is labeled as PT, where P is the parity operator and T is the time-reversal operator; without spin-orbit coupling this is simply a spin-flip operator. Furthermore, because the crystal has inversion symmetry, the pair amplitude (or order parameter) separates into even and odd parity terms, as shown in Equation 1. Because of fermion antisymmetry, even-parity states are pseudo-spin singlets and odd-parity states are pseudo-spin triplets, unless the pair state is odd in frequency, where one then has even-parity triplets and odd-parity singlets. One can then expand the order parameter in terms of basis functions for each irreducible representation of the crystal point group. This is discussed in detail by Yip and Garg [54], who give the basis functions for the cubic and hexagonal crystal types for strong spin-orbit coupling. In a hexagonal lattice at the Brillouin zone center (corresponding to zero center-of-mass pairing) there are four one-dimensional ($A_1$, $A_2$, $B_1$, $B_2$) and two two-dimensional ($E_1$, $E_2$) representations for each parity state, with the parity labeled as *g* for even and *u* for odd. For convenience, we give some of the basis functions for hexagonal symmetry in Table IV. Note that these are expressed in powers of the vector **k**, but could as well be expressed as spherical harmonics $Y_{LM}(k)$. In hexagonal symmetry these can mix with one another for *L* values differing by 2 and *M* values differing by 6. An unconventional superconductor would be any representation besides $A_{1g}$. Note that this has nothing to do with nodes in the gap, because an $A_{1g}$ pairing state can have zeros in the gap. (For example, $Y_{20}$, which has two line nodes, transforms as $A_{1g}$ in hexagonal symmetry.) Consider also a d-wave pairing state. The basis functions for this state have two powers of **k**, corresponding to $L = 2$, as seen in Table IV. Thus the d-wave state decomposes into an $A_{1g}$ state ($Y_{20}$), an $E_{1g}$ state ($Y_{21}$, $Y_{2-1}$) and an $E_{2g}$ state ($Y_{22}$, $Y_{2-2}$). Furthermore, the $A_{1g}$ state can mix with other states ($Y_{00}$, $Y_{40}$, etc.), and thus a d-wave pairing state may or may not be unconventional.

The order parameter η for multi-dimensional representations is often written as a complex vector, with each component associated with a particular basis function. For example, in the two-dimensional $E_{2u}$ case shown in Table IV one can have

$$\Delta(\mathbf{k}) = \eta_1 k_z(k_x^2 - k_y^2)\mathbf{z} + \eta_2 k_z(2k_x k_y)\mathbf{z} \quad \text{with } \eta = (\eta_1, \eta_2). \tag{2}$$

For the case $\eta = (1, i)$ one would obtain $k_z(k_x + ik_y)^2\mathbf{z}$, where $\mathbf{z}$ denotes the spin vector for $S = 1$ and $M_S = 0$. This combination is obtained for weak coupling BCS theory since it minimizes the phase space of nodes (it has point nodes plus a line node). Such an order parameter breaks time-reversal symmetry, however. The actual superconducting state, assuming it belongs to this representation, could be (1,0) or (0,1), however, which would then have additional line nodes. Such states could be stablized either by the presence of a symmetry breaking field (which could prefer one orientation or the other) or by strong coupling effects (such as occur in $^3$He).

We note that the form of the order parameter discussed by Yip and Garg and shown in Table IV is only valid near $\mathbf{k} = 0$. A proper order parameter has the property that it is invariant under reciprocal lattice translations. One can generate such basis functions by a tight binding expansion: $f_\Gamma(\mathbf{k}) = \Sigma_\alpha c_\alpha^\Gamma e^{i\mathbf{k}\cdot\alpha\mathbf{R}}$, where $\mathbf{R}$ is a lattice vector, $\alpha$ a group operation, and $c_\alpha^\Gamma$ is the character of the representation $\Gamma$ for operation $\alpha$ (a number for single-dimensional representations, a matrix for multi-dimensional ones). In general, the order parameter is a linear combination of these $f_\Gamma(\mathbf{k})$ for different lattice vectors, but usually one vector will be dominant since the pair interaction is local in space in most models. An interesting point is that the nodes of these functions can differ from those of Table IV for Fermi surfaces not centered at k=0. This has been demonstrated in hexagonal symmetry by Putikka and Joynt [55]. They also considered functions generated by non-primitive translation vectors (connecting the two uranium atoms in the hexagonal close-packed cell), which are not allowed because they are not invariant under reciprocal-lattice translations. This technical point has caused confusion, and will be discussed further in the next section.

Another point which has caused some confusion comes up for odd parity, where Blount has shown that line nodes are not allowed for strong spin-orbit coupling [56]. This does not hold if the order parameter contains less than 3 pseudo-spin components, however. Also, even if all 3 pseudo-spin components are present, in special cases one can have line nodes due to the fact that the Cooper pair is a product of single-particle wavefunctions at $\mathbf{k}$ and $-\mathbf{k}$, and certain $\mathbf{k}$ points have special transformation properties. An example of this will be discussed later.

In recent years, most theoretical attention has been focused on UPt$_3$ because its unusual and well-studied phase diagram gives important clues to the symmetry of the order parameter. We therefore emphasize UPt$_3$ in order to illustrate the modern theoretical approaches to HF superconductivity. To begin with, the split transition in zero field indicates that some degeneracy is being lifted. The three degeneracy categories that have been proposed are orbital, spin, and center-of-mass as summarized in Table V. We discuss each in turn.

The most commonly explored broken symmetry is the point group symmetry, and the most popular model assumes that the split transition is due to lifting of orbital degeneracy in the pair wavefunction. Most models assume that the transition is due to one of the two-dimensional representations, E$_1$ or E$_2$ [57]. The degeneracy of this state could then be broken by the weak antiferromagnetism since it has a lower symmetry than hexagonal (orthorhombic), with the T$_c$ splitting predicted to be proportional to the moment squared, just as observed in the neutron scattering experiments [42]. Such splitting would correspond to a condensation of one component of the order parameter η at the upper transition. This can be seen by considering the two-component order parameter (η$_1$, η$_2$) as a vector in the basal plane, which is oriented by the magnetic-moment vector either parallel or perpendicular to the b-axis, depending on the sign of the coupling. At lower temperatures, the superconducting condensation energy wins out over the orientation energy and the second component of the order parameter becomes non-zero. This produces another jump in the specific heat, as well as an increase in the lower critical field (thus giving a kink). Both effects have been observed experimentally. Similar effects have been also observed in Th-doped UBe$_{13}$. Thus the order parameter of phase A can be labeled as (1,0) and phase B as (1,i) (See Fig. 2). In weak coupling theory, the (1,i) phase relation is the most stable. Now imagine we apply a field **H** in the basal plane. If the magnetic anistropy is weak (which is inferred experimentally from magnetoresistance measurements [58]), the Zeeman energy will rotate the AFM moment **M** so that it is perpendicular to **H**. Thus, at a high enough field, the orientation effect of **H** will exceed **M** and the superconducting state will flip from being (1,0) to being (0,1). This will cause a kink in the upper critical field as observed. The B-C phase line will then correspond to going from (1,i) to (0,1).

A problem, though, arises for fields out of the basal plane. A field along the c-axis does not break hexagonal symmetry and so causes no orientation effect. Thus, no tetracritical point is predicted for this direction (phase lines of the same symmetry cannot cross), in complete contradiction to experiment. Theoretically, this arises from gradient terms in the Ginzburg-Landau (GL) theory which mix the two order-parameter components, causing a repulsion of the phase boundaries. Within the context of the $E_1$ or $E_2$ models, several solutions have been proposed. For fields along the c-axis, the solution near $H_{c2}$ can be of the form (a,bi) [59]. Because of this, a B-C phase line may still exist between (1,i) and (a,bi), as suggested by Vorenkamp [58]. On the other hand, Joynt pointed out some time ago that in this case the A phase (1,0) continuously develops into the C phase (a,bi), and so no tetracritical point exists [59]. A second proposal by Joynt et al [60] was that a superconducting glass phase could occur, due to the fact that the superconducting order parameter might not be single-domain since the coherence length of the antiferromagnetism is comparable to that of the superconducting phase. Thus the superconducting order parameter in the A phase could point in 1 of 6 equivalent directions. This theory can give a tetracritical point for a field along the c-axis, but at the expense of turning the A-B phase line into a first-order one, which does not appear to be consistent with current data. Also, the weak anisotropy of the AFM moment would indicate that only a small field is needed to cause domain orientation, which would also act to spoil this argument.

A third proposal is due to Sauls [61]. He noted that for an axially symmetric Fermi surface, the gradient term in the GL theory, which mixes the two order parameter components and destroys the tetracritical point for fields along the c-axis, vanishes for the $E_2$ representation. Hexagonal anisotropy spoils this argument, however. Sauls has therefore argued that the lack of basal-plane anisotropy in the upper critical field [62] is in support of the conjecture that the effective hexagonal anisotropy must be small. On the other hand, theoretical calculations based on the actual Fermi surface of $UPt_3$ (which has strong hexagonal anisotropy, as seen in deHaas-vanAlphen experiments [63]) indicate that a very small value of this mixing term would either have to be due to cancellation between various Fermi surface sheets (and so would have to be somewhat coincidental) or that the effective order parameter and Fermi surface has axial symmetry for some unknown reason [64]. A small ratio of the mixing term to the main gradient term (10%), though, is consistent with the calculations, and may also be consistent with the data because the splitting of the tetracritical point can be shown to go as the square of this ratio [64].

The second part of Sauls' argument concerns the observation that the upper critical field lines for fields in the basal plane and for fields along the c-axis cross at low temperatures [62]. The data have been shown to be consistent with a suppression of $H_{c2}$ for fields along the c-axis due to Zeeman splitting [65], a phenomenon known as Pauli limiting. Since the observed susceptibility is actually larger for fields in the basal plane, the largest Pauli limiting effect would happen for fields in the basal plane if $UPt_3$ was a spin singlet superconductor. This contradicts experiment, however. The only remaining possibility is that $UPt_3$ is a spin-triplet superconductor, but with only one component condensed: the $d_z$ one. This state is just $M_S = 0$ relative to the c-axis; that is, the Cooper pair spins are confined to the basal plane. Because of this, such a state would suffer no Pauli limiting for fields in the basal plane (since the Cooper pair spin can align with the field), but would be Pauli limited for fields along the c-axis, as experimentally observed. Such a state can arise in spin-fluctuation models for the pairing because the spin fluctuations are experimentally known to be confined to the basal plane in $UPt_3$. This state is taken to be a signature of strong spin-orbit coupling because the spin vector is locked to the lattice. In fact, spin-orbit coupling would cause this state to mix with the other two spin components (that is, the locking really applies to J, not S). Because of this, the model at hand is a hybrid, keeping part of the spin-orbit effect (locking to the lattice) and ignoring the other (spin-component mixing). Finally, we should comment that the pairing state of this model has $E_{2u}$ symmetry, and would thus appear first for f-wave pairing ($Y_{32}$ harmonic). The gap function has line nodes perpendicular to the c-axis and point nodes along c, consistent with anisotropy seen in transverse ultrasound data [17] and in the penetration depth [16]. Moreover, the gap near the point nodes has a quadratic dispersion (as opposed to the linear dispersion in the $E_{1g}$ case), which is almost certainly necessary to explain recent low temperature thermal conductivity data in $UPt_3$ [66]. This difference between $E_{2u}$ and $E_{1g}$ is illustrated in Fig. 10.

The failure of the E model to predict a tetracritical point for fields along the c-axis has led to another proposed orbital scheme, two nearly degenerate representations, taken to be either 2 one-dimensional representations [60,67] or a one-dimensional and a two-dimensional one [60,68]. For this model, a tetracritical point is allowed for all field orientations since the two representations are different; that is their upper critical field lines correspond to different quantum numbers and thus are allowed to cross [69]. The main problem with this idea is that there is no version of this model which couples the splitting of $T_C$ to the antiferromagnetism, because the $T_C$ of the two representations are already split in hexagonal symmetry and so the weak moment should play no role. Furthermore, one would expect the two $T_C$ lines in this model to cross under pressure, not to merge. (Recent pressure experiments by Boukhny et al [13] claim a crossing, consistent with the two representation model; others by Sieck et al [13] claim a merging, consistent with the E model). Finally, a small splitting in $T_C$ between two different representations implies that the system has almost spherical symmetry, since $T_C$ depends exponentially on the coupling constant (this is the basis for the model of Ref. 68). Such a model is therefore more restrictive than the above-mentioned $E_2$ model, which only requires approximate axial symmetry.

The models described above concern a lifting of orbital degeneracy. An alternate model has been proposed by Machida and co-workers based on a lifting of spin degeneracy [70]. This assumes weak spin-orbit coupling, which would seem to be totally contradictory because the single particle states are known to be in the strong spin-orbit limit. In their model, the pairing state is an orbital singlet (taken to be $A_{1u}$) and a spin triplet. Because the spins are not tied to the lattice by spin-orbit coupling, there is no problem in obtaining a tetracritical point for all field orientations. Nevertheless, there are several problems with this model. First, because the order parameter has 3 components, the theory predicts more superconducting phases than observed experimentally. The authors have attempted to resolve this problem by splitting off the $M_S = 0$ component of the triplet by spin-orbit coupling to crystal fields. The theory then reduces to the E model discussed earlier, but with a double degeneracy in the spin part of the wavefunction. The problem, though, is that Choi and Sauls have shown that such a state ($M_S = \pm 1$) is inconsistent with the observed anisotropic Pauli limiting of the upper critical field discussed earlier [65]. In the latest version of the Machida model, it is pointed out that the theory predicts gapless behavior for one of the two spin states. This yields a residual density of states $\gamma_r(0)$ about half the normal state value, which is seen for some UPt$_3$ samples. The latest specific heat work of the Grenoble group [71], however, indicates a very small residual $\gamma_r(0)$ component, reinforcing a belief that the residual component seen in some samples is associated with impurity pair breaking, as opposed to intrinsic behavior. Finally, it should be noted that the argument for gaplessness does not work if all 3 components of the spin triplet are involved in the gap function.

A third type of degeneracy in the order parameter is that associated with center-of-mass momentum. In standard BCS theory, the electrons pair with zero center-of-mass momentum (**k**,-**k**). In a recent microscopic theory for heavy fermion superconductivity proposed by Coleman and associates [72], the pair wavefunction is odd in frequency. One consequence of this is that such a pair state must be non-uniform, even at zero field, due to its negative Meissner stiffness. This means that the pairs want to have a finite center of mass momentum (**k**,-**k**+**Q**). Recently, Heid et al [73] have exploited this fact to propose yet another model for the phase diagram of UPt$_3$. In their model, they take **Q** to be the M point of the zone. This vector is one of three inequivalent vectors (that is, the M point has a star of 3). Thus, the pair state has 3 components whose degeneracy can again be broken by the antiferromagnetism. Moreover, this model can have a tetracritical point for all field directions, and the resulting Ginzburg-Landau theory has rich behavior. One problem, though, is that for some choices of the G-L parameters in the free energy, all 3 components of the order parameter are non-zero, yielding too many superconducting phases, as discussed above for the Machida model. These problems could perhaps be avoided by pairing at the K point, because its star is 2 instead of 3.

It should also be remarked that pairing at a finite center-of-mass momentum can occur in another model as well. Imagine that one has local antiferromagnetic order, and that one pairs states which are diagonal in this order. Then when one transforms back to a global hexagonal symmetry, a part of the pair state corresponds to the pairing (**k**,-**k**+**Q**), where **Q** is the antiferromagnetic wavevector. For UPt$_3$, **Q** is at the M point, so such a model would look similar to the one of Heid et al, even though the physics is of different origin. This type of theory was worked out for the case of magnetic superconductors, such as the Chevrel phase materials [74]. There are some theories of heavy fermion superconductivity based on these ideas [75], but they do not consider the multi-component nature of the order parameter. Such an application could turn out to be of potential use in explaining UPt$_3$. On the other hand, it should be remarked that it remains to be shown whether the order parameter component associated with Q can survive in a self-consistent treatment due to phase space considerations (that is, $\varepsilon_k$ and $\varepsilon_{-k+Q}$ are not in general degenerate, which in turn reduces the pair susceptibility).

The case of $U_{1-x}Th_xBe_{13}$ is also very interesting. Note in Figure 6 that three phase lines meet at both $x_c \approx 0.019$ and $x_c \approx 0.043$ doping. As pointed out by Luk'yanchuk and Mineev [76], such tricritical points are not allowed if all the phase lines are second order, which they appear to be. This implies the existence of other phase lines, as indicated by the dashed lines in Figure 6. Since these lines are vertical as a function of doping, they are hard to access experimentally. As discussed previously, this difficulty was bypassed by noting that pressure behaves much like doping, so the phase line near $x_c \approx 0.019$ was searched for by looking at the specific heat as pressure was varied [28]. From the different values of $dT_c/dP$ for $x > x_c$ and $x < x_c$ (both at $P = 0$) and a model of two different superconducting states which cross at $x = x_c$, Sigrist and Rice [77] found theoretically that $x_c = 0.018 + 0.0017$ P(kbars). Indeed, C/T jumps at the predicted pressure, thus confirming the vertical nature of the phase line suggested from Figure 6. So far, this has only been done for the 1.9% phase line; the 4.3% one remains to be looked at. Finally, in pure $UBe_{13}$, a second phase line in the H-T phase plane has been inferred from specific heat data and occurs at about 2 Tesla with a weak temperature dependence [78], very similar to what was observed in $UPt_3$.

The basic model attempting to explain this behavior was put forth by Kumar and Wolfle [79]. The pure phase is taken to be d-wave. This phase is suppressed as a function of Th doping due to pair breaking by the Th impurities. It is assumed that there is an s-wave phase with a smaller value of the critical temperature. Because an s-wave phase is less sensitive to non-magnetic impurities than a d-wave one, the two phase lines will cross at a critical doping, leading to a pure s-wave phase, followed at lower temperatures by a mixed s-d phase. This mixed phase is predicted to be of the form s + id, so it breaks time reversal symmetry, consistent with the μSR experiments. Moreover, the gap would be nodeless, in agreement with recent specific heat data in the second phase [26], and in contrast to specific heat data in the pure phase which indicate the presence of point nodes in the gap (as would occur for a pure d-wave state in cubic symmetry). This was discussed above in Section IV B. (It should be mentioned again that a large residual specific heat term is seen in the Th-doped phases, the origin of which is still being debated).

We note that this s-d model for (U, Th)Be$_{13}$ is the same as the nearly degenerate representation model discussed previously for UPt$_3$. Sigrist and Rice [77] have analyzed this model in great detail for a variety of different sets of assumed representations and showed that the unusual doping dependence of the phase lines could be explained by such models. Moreover, they thoroughly discussed jumps in C/T and H$_{c1}$, and the analysis of ultrasound attentuation and μSR data in the context of these models. The data put strong constraints on the possible representations which are allowed. Sigrist and Rice suggest that the two phases are actually a one-component and a multi-component p-wave state, with the non-unitary nature (**dxd**\* non-zero) of the latter responsible for the weak magnetism. Further support for an odd-parity state has been given by Han et al [80], who observed a negative proximity effect between UBe$_{13}$ and Ta (an s-wave superconductor).

Finally, we should remark that there is no clear evidence of a multi-component order parameter in the other heavy fermion superconductors. This does not mean, though, that the superconductivity of these other metals is not unconventional (there are, after all, more non-trivial 1D representations than 2D and 3D ones). For example, the split transitions observed in UPt$_3$ would not be expected to be observed in other heavy fermion superconductors, even if their pair state was multi-dimensional. In URu$_2$Si$_2$ this is because the moments point along the c-axis and so do not lower the symmetry. In UPd$_2$Al$_3$ and UNi$_2$Al$_3$ the moments are so large that any predicted T$_c$ splitting should be of the order of T$_c$ itself, so that only one transition would be seen. Nevertheless, many of the thermoydnamic measurements done for UPd$_2$Al$_3$, UNi$_2$Al$_3$, and URu$_2$Si$_2$ produce similar results to UPt$_3$, and thus are indications that the gap function may have line nodes in these cases too. The problem is that since line-node states can also occur in the trivial A$_{1g}$ representation, it will be much more difficult to ascertain whether the pair state is unconventional in these cases. Similar problems also exist in the area of high temperature superconductivity, and is an area of great current interest there as well.

## VII. Pairing Mechanisms

The pairing mechanism in heavy fermion superconductors is one of the most difficult problems in condensed matter physics today. This is evident from a comparison with $^3$He. Heavy fermion superconductors possess complicated crystal structures, orbital degeneracy of the f-electrons, hybridization with non-f electrons, and strong spin-orbit effects. Also, there is the additional problem of determining whether the pairing mechanism is solely electronic in nature or of electron-phonon origin. If it is a combination, then the physics will certainly be extremely difficult. In discussing these issues, we first give some arguments against an electron-phonon mechanism, then turn to a discussion of various possible electronic mechanisms. We note that some of these issues have been covered in an earlier review in this journal [81], which is recommended reading for anyone interested in this field.

Pairing can occur for a state that avoids the repulsive effects of the Coulomb interaction. In the electron-phonon mechanism, the pairs are correlated in time, being at the same place (to take advantage of the attraction due to the phonons) but at different times (to avoid the Coulomb repulsion). This works in the electron-phonon case because the time-scale of the phonons is much longer that that of the electrons. For heavy fermions, however, the Coulomb repulsion between the f-electrons is so strong that it is difficult to get a net attractive electron-phonon interaction. This is exacerbated by the fact that the Fermi energy is actually smaller than the Debye frequency, and therefore one is in the opposite limit from ordinary superconductors. One possible way out is that for phonon frequencies larger than the Fermi energy, the electrons are unrenormalized; that is, the electron-phonon interaction may look normal in this frequency regime and so one can have phonon-mediated pairing. Strong-coupling calculations show this not to be the case, though. The mass renormalization effects associated with the heavy fermions still act to renormalize the electron-phonon interaction to a very small number, and one predicts very small values of $T_C$ even if the Coulomb repulsion is totally ignored [82]. This comes from the simple idea that the effective coupling constant is of the form $\lambda_{ep}/(1 + \lambda_{hf})$, where $\lambda_{ep}$ is the electron-phonon coupling and $1 + \lambda_{hf}$ is the heavy fermion mass renormalization. Calculations of $\lambda_{ep}$ based on the rigid-muffin-tin approximation give very small values indeed (of order 0.1) [83], so that $\lambda_{hf} \gg \lambda_{ep}$, and therefore the effective coupling constant is very small. It is possible that because the heavy fermion mass renormalization is very sensitive to volume (large Gruneisen parameter), there could be a large indirect electron-phonon effect which might lead to pairing. This same mechanism is thought to be connected to the α-γ phase transition in cerium, where it is known as the Kondo volume-collapse model, and is the basis of a pairing theory put forth by Fulde's group [84]. Current calculations with this model, though, indicate that the Coulomb repulsion still wins out with no net attraction [85]. This leaves consideration of inter-site pairing due to phonons, early theories of which were reviewed by Lee et al [81]. However, no rigorous calculations, such as that done for the on-site case [85], exist at this time. It should be remarked that although phonon softening effects have been seen in the superconductor $U_6Fe$, such effects have not been seen in any heavy fermion superconductor to date [86]. This fact tends to cast additional doubt on the future success of any electron-phonon model for pairing in heavy fermions.

A prime argument in favor of an electronic mechanism for the HF pairing is the similarity with liquid $^3He$. For example, the f-electrons in the HF metals are both nearly localized and nearly magnetic, just as the helium atoms are in $^3He$, which is a superfluid with p-wave pairing. Even though a completely quantitative microscropic theory of the pairing in $^3He$ is lacking, most known properties can be explained on the basis of spin-fluctuaton mediated pairing. This idea has been explained most simply by Leggett [87]. In $^3He$ the pairs must be in a higher angular momentum state ($L > 0$) in order to avoid the hard core repulsion of the helium atoms. This is insured for an $S = 1$ state since fermion antisymmetry dictates that $L$ must be odd. Why then does pairing occur? The answer is basically that if one atom has spin up, its Zeeman field lowers the energy of the second atom if it also has spin up. Atom one in turn has its energy lowered by the second atom's Zeeman field, and this cooperative effect leads to pairing. An $L = 1$ state is preferred, as observed experimentally, since the interaction becomes weaker in higher $L$ channels.

Spin fluctuations can also explain another important fact in $^3He$. Surprisingly, under pressure the pair state switches from the so-called B phase, which is isotropic, to an anisotropic A-phase. Anderson and Brinkman [88] showed that a spin-fluctuation model could explain this, because in the anisotropic A phase the relative pairing interaction is enhanced below $T_C$ due to an anisotropic reduction of the spin susceptibility. In the B phase, this reduction is isotropic and leads to a relative decrease in the pairing interaction. In reality, the situation is more complicated than this. The Landau scattering parameters inferred from experiment indicate that other interactions, such as transverse-current interactions, also play an important role [89]. In addition, the bare interaction in $^3He$ is attractive at intermediate distances due to van der Waals forces. These are ignored in spin-fluctuation models, but play a role in polarization-potential models [90]. Given these facts, it is somewhat amazing that strong-coupling spin-fluctuation models work as well as they do in describing the pairing properties of $^3$He. Nevertheless, after the discovery of heavy fermions, the close correspondence of these metals to $^3$He (large mass renormalization, nearness to magnetism and localization) was noted [91], and attempts were made to apply $^3$He phenomenology to heavy fermions [81]. However, because $^3$He is an isotropic system with a single band and a spherical Fermi surface, while HFs have multi-sheet Fermi surfaces, orbital degeneracy, crystal fields, and spin-orbit effects, these attempts were largely unsuccessful.

Stimulated by the discovery of antiferromagnetic spin fluctuations in UPt$_3$ [5], it was shown theoretically that such fluctuations would lead to $L = 2$, $S = 0$ pairing, compared to ferromagnetic spin fluctuations, which lead to $L = 1$, $S = 1$ pairing [92]. How does this work? In an $L = 2$ state, the pair wavefunction is largest when the electrons are on near-neighbor sites, thus reducing the Coulomb repulsion. Now assume that they have a singlet-spin orientation. Then the antiferromagnetic spin interaction between the two electrons will cooperatively lower each other's energy similar to the ferromagnetic case of $^3$He described above. In this argument, it is assumed that the antiferromagnetic **Q** vector corresponds to near neighbors with antiparallel spins. Thus, one has d-wave pairing. This model has now been applied to high $T_C$ cuprates and is one of the leading models for describing their pairing.

The deficiencies of these models were soon realized, however. They were developed for a single band metal with simple cubic geometry. In reality, all known HF superconductors have at least two f-atoms in their low-temperature unit cell. The most commonly studied examples of UPt$_3$ and UBe$_{13}$ have non-symmorphic lattices; that is, the two f-atoms in the unit cell are separated by a non-primitive translation vector. This means that the problem cannot be reduced to a single-band form. Trying to do so leads to gap functions which are not properly invariant under translation by a reciprocal lattice vector [93,55]. There have been many attempts to get around this problem, none of which have been completely satisfactory [94]. The latest attempt involves using wavefunctions from band structure calculations, thus taking into account the proper phasing relation between the two f-atoms of the unit cell [95]. These calculations almost always yield "anisotropic s-wave" gaps, that is, the gap is from the identity representation $A_{1g}$. (Note, in hexagonal symmetry, $Y_{20}$, $Y_{40}$, etc., transform as $A_{1g}$ so the notation

"anisotropic s-wave" can be misleading).  A simple way to see this comes from the earlier results of Norman [93,82].  The antiferromagnetic Q vector for the high frequency spin fluctuations in $UPt_3$ corresponds to a reciprocal lattice vector of $(0,0,2\pi/c)$.  In a double zone scheme, this Q vector connects the bonding and antibonding Fermi surfaces.  (In real space, it connects the two uranium planes in the unit cell.)  The solution of the gap equation produces an anisotropic s-wave gap of one sign on one surface and the opposite sign on the other surface; this sign change plays the same role as the sign change of a d-wave gap under point group operations.  It is amusing to note that this s,-s model has been rediscovered as a possible model for two-layer high $T_c$ cuprates like YBCO.  To summarize then, the early arguments leading to pure d-wave pairing in heavy fermions are only specific to single-band models with a simple Fermi surface and almost certainly do not apply to real HF metals.

Recently, one of the present authors has considered a more sophisticated spin fluctuation model [96].  This model replaces the simple contact interaction used in earlier spin fluctuation models (which only depends on the relative spin orientation) by the screened Hartree-Fock interaction between the f-electrons.  This therefore includes orbital and spin-orbital interaction effects not present in $^3$He.  Surprisingly, this model yields a solution even for on-site pairing.  The pairing state has a relative $J = 4$, where $J$ is the total angular momentum.  This state is largely $L = 5$, $S = 1$ in nature, possessing the maximal allowable $J$ for two $j = 5/2$ f-electrons on the same site, which minimizes the Coulomb repulsion between the two electrons.  That is, this pairing state is the low temperature quasiparticle analogue of the high temperature Hunds-rule ground state of an $f^2$ ion.  The degeneracy of the state is broken by the crystal environment.  This effect is included by projecting all $J = 4$ states on the Fermi surface and finding which one has the largest projection, thus determining the representation with maximum $T_C$.  An $E_{2u}$ state was found for $UPt_3$, which is one of the phenomenological models discussed in the last section (Other representations have similar $T_c$'s, though, so a nearly degenerate representations model is also possible.) Note that an odd-parity state in this case corresponds to an order parameter which changes sign from one uranium site in the unit cell to the other.  A model of this type was first proposed by Anderson [97].  Such a state has a higher $T_C$ than one of even parity since 3 pseudo-spin components contribute (as opposed to 1 component for the even parity state), thus giving a larger projection on the Fermi surface.  We point out that there is no relation between the $S = 1$ nature of the $J = 4$ state and the fact that the order parameter is a pseudo-spin triplet. This point is mentioned to reinforce the fact that analogies from $^3$He case can sometimes be very misleading when applied to HFs. Because of spin-orbit coupling, all 3 components of the pseudo-spin triplet occur in the gap function at a general **k** point. Despite this, the gap function identically vanishes in the $k_z = \pi/c$ zone face; that is, one has line nodes. This surprising result is due to the special transformation properties of **k** points on the zone boundary face in a non-symmorphic lattice. Similar considerations for the case of near-neighbor pairing due to antiferromagnetic correlations give a state with relative $J = 0$, corresponding to the $S = 0$ state discussed in the spin only case above. Although this does not necessarily imply an identity representation ($A_{1g}$) for the gap function, calculations do show such a representation as having the largest $T_C$ [96]. We have gone into some detail about this model to illustrate how drastically one's conclusions change when considering real f-electrons in a crystal environment, as opposed to an isotropic single band system.

A variety of other electronic pairing models have been proposed in the literature besides spin fluctuation ones. A popular model involves charge fluctuations in a slave boson approach leading to a weak d-wave pairing instability. In this case, the pairing boson is a fluctuation between f and conduction electrons [98]. This model has only been applied to simple single band systems, so how it will look in a more realistic system is unknown. More recently, Cox has proposed a pairing mechanism based on the quadrupolar Kondo effect, which can arise in the multi-channel Kondo problem [99]. Strong constraints on the possible pairing state in UPt$_3$ are predicted by this model, but under very restrictive assumptions concerning the assumed crystal-field ground state of the f-ion and the available conduction-electron partial wave channels. A pairing state even in frequency is assumed.

As it turns out, odd-frequency pairing states are possible in the multi-channel Kondo model, too. By exploiting this, Heid et al [73] have proposed a new model for UPt$_3$ based on finite center-of-mass momentum pairing in the context of a two-channel Kondo model discussed in Section VI. This work was motivated by a recent theory of Coleman *et al.* [72], who propose that HF superconductivity is due to a peculiar 3-particle condensate which decomposes into a Cooper pair and a charge-neutral singlet. The Cooper pairing is odd in frequency and can have a finite center-of-mass momentum, as discussed in the last section. This model yields a large residual specific heat coefficient $\gamma_r(0)$ in the superconducting state, due to the fact that the odd-frequency gap function vanishes on the Fermi surface. The authors argue that the absence of an increased low-temperature NMR relaxation rate in UPt$_3$ is strong support for their model because a large $\gamma_r(0)$ due to impurities would induce a residual (Korringa) component in the NMR relaxation rate, which is not observed. On the other hand, recent thermal conductivity data in UPt$_3$ [66] do not see a residual $\gamma_r(0)$ component, which this theory would predict to occur.

Other models have also been proposed. A recent one involves a conventional pairing state, but takes into account local antiferromagnetic order which causes a line node in the excitation gap perpendicular to the antiferromagnetic **Q** vector [75]. This model was developed earlier to explain certain properties of the magnetic Chevrel phase superconductors. A drawback to such a model is that the argument producing a line node applies only to a special Fermi surface, and thus should break down for the more complicated Fermi surfaces one actually sees in HF metals. The model also does not address the question of a multi-component order parameter.

How will the issue of the pairing mechanism be resolved in heavy fermion metals? Here is a possible scenario. UPt$_3$ currently provides us with the most information, given its rich phase diagram. There is therefore hope that additional experimental work and the appropriate phenomenological theory will yield the order parameter in a relatively short amount of time. Such a model could then put very strong constraints on a possible microscopic theory. More information, though, may be necessary. One way to proceed is to obtain additional detailed quasiparticle mass information from deHaas-van Alphen measurements. By comparing this to an adjusted band model which agrees with the expermental Fermi surface topology, one can extract the momentum dependence of the mass renormalization, which then can be fit to a self energy function. From this, one should be able to construct an analogous model for the self-energy in the particle-particle channel, and then proceed to find out which pair state should have the highest $T_C$. Once this has been achieved, the rest of the heavy fermion superconductors can be studied to see whether such a theory also predicts their properties. Since we now have many examples of HF superconductors, another route may be to look at trends in the HF metals. An analysis of such trends led to one of the spin fluctuation models discussed above [96]. There, it was the observation that all but one of the HF superconductors were uranium alloys and that the available evidence points to an $f^2$ ground state for the uranium ion. Perhaps new dramatic insights in the future (as well as the discovery of other HF superconductors) will lead more quickly to the correct microscopic theory of these amazing superconductors.

Acknowledgment: Work at Los Alamos was carried out under the auspices of the U. S. Department of Energy. Work at Argonne was supported by the U. S. Department of Energy, Basic Energy Sciences, under Contract No. W-31-109-ENG-38.

**Table I**

| Material | $\gamma(T_c)$ (J/mol-K$^2$) | $m^*/m$ | $\rho(T_c)$ ($\mu\Omega$ cm) | $\rho(300K)$ ($\mu\Omega$ cm) | $T_{max}$ (K) | $\chi(T_c)$ ($10^{-3}$ emu/mol) | $T_N$ (K) | $\mu$ ($\mu_B$) |
|---|---|---|---|---|---|---|---|---|
| CeCu$_2$Si$_2$ | 0.73-1.1 | $\leq$ 380 | 2-65 | 50 | ~ 18 | 80 | 1.3 | .1-.3 |
| URu$_2$Si$_2$ | 0.065 | 140 | 12-70 | 330/170 | ~ 60 | 1.2/4.9 | 17 | 0.03 |
| UPd$_2$Al$_3$ | 0.145 | 66 | 4 | 150/200 | ~85 | 110/30 | 14 | 0.85 |
| UNi$_2$Al$_3$ | 0.120 | 48 | 7 | 1000 | ~200 | 45/30 | 4.6 | 0.24 |
| UPt$_3$ | 0.450 | 180 | 0.3-3 | 230/135 | >300 | 110/60 | 5.0 | 0.02 |
| UBe$_{13}$ | 1.100 | 260 | 18 | 110 | ~2 | 15 | - | - |

Table caption: Typical normal state parameters for the six known ambient pressure HF superconductors. Values are taken from the references quoted in the text. Values separated by a slash (/) indicate a-axis/c-axis anisotropies. Dashes indicate a range of measured values.

**Table II**

| Measurement | Polar (lines) | Axial (points) |
|---|---|---|
| Specific Heat (C) | $T^2$ | $T^3$ |
| NMR Relaxation ($1/T_1$) | $T^3$ | $T^5$ |
| Thermal Conductivity ($\kappa$) | $T^2$ | $T^3$ |
| Penetration Depth ($1/\lambda_\parallel^2$) | $T^3$ | $T^2$ |
| Penetration Depth ($1/\lambda_\perp^2$) | $T$ | $T^4$ |

Table Caption: Theoretical temperature dependencies for several low temperature measurements, assuming a spherical Fermi surface and either line or point nodes in the superconducting gap structure.

**Table III**

| Material | $T_c$ (K) | $\lambda$ (1000 Å) | $\xi$ (Å) | $H_{c1}(0)$ (mT) | $H_{c2}(0)$ (T) | $H'_{c2}(T_c)$ (T/K) | $\Delta C/(\gamma(T_c)T_c)$ |
|---|---|---|---|---|---|---|---|
| $CeCu_2Si_2$ | 0.7 | >4 | 90 | 2.3 | 2.0/2.4 | -23 | 1.4 |
| $URu_2Si_2$ | 1.2 | >15/9–10 | 100-150 | 1.4 | 14/3 | -4.0 | 0.9 |
| $UPd_2Al_2$ | 2.0 | 5/5 | 85 | 1.0 | 3.0/3.6 | -4.3 | 1.1 |
| $UNi_2Al_2$ | 1.0 | >3 | 240 | 1.5 | 1.5 | -1.4 | $\geq 0.5$ |
| $UPt_3$ | 0.55 | 6-7/6-7 | 100-120 | 3.0 | 2.8/2.1 | -4.5/-7.7 | .6 -1.0 |
| $UBe_{13}$ | 0.9 | >8 | 100 | 4.6 | 10.1 | -44 | 2.0 - 2.6 |
| Al | 1.2 | 0.05 | 16000 | - | 0.01 | | 1.4 |
| Nb | 9.5 | 0.039 | 380 | - | 0.21 | | 2.0 |
| $V_3Si$ | 16 | 0.062 | 30-50 | 78 | 18-20 | -2.0 | 2.0 |

Table Caption: Selected superconducting parameters for the HF and other superconductors. Values are taken from the references quoted in the text. Values separated by a slash (/) indicate a-axis/c-axis anisotropies. Dashes indicate a range of measured values.

**Table IV**

| Dimension | Representation $\Gamma$ | Basis Functions | Representation $\Gamma$ | Basis Functions |
|---|---|---|---|---|
| 1 | $A_{1g}$ | $1, k_x^2 + k_y^2, k_z^2$ | $A_{1u}$ | $k_z \mathbf{z}$; $k_x \mathbf{x} + k_y \mathbf{y}$ |
| 1 | $A_{2g}$ | $\mathrm{Im}\, k_+^6$ | $A_{2u}$ | $\mathrm{Im}\, k_- \mathbf{r_+}$; $k_+^6 k_z \mathbf{z}$ |
| 1 | $B_{1g}$ | $k_z \,\mathrm{Im}\, k_+^3$ | $B_{1u}$ | $\mathrm{Im}\, k_+^3 \mathbf{z}$; $k_+^2 k_z \mathbf{r_+}$ |
| 1 | $B_{2g}$ | $k_z \,\mathrm{Re}\, k_+^3$ | $B_{2u}$ | $\mathrm{Re}\, k_+^3 \mathbf{z}$; $k_+^2 k_z \mathbf{r_+}$ |
| 2 | $E_{1g}$ | $k_z k_+$; $k_z k_-^5$ | $E_{1u}$ | $k_+ \mathbf{z}$; $k_z \mathbf{r_+}$ |
| 2 | $E_{2g}$ | $k_+^2$; $k_-^4$ | $E_{2u}$ | $k_+ \mathbf{r_+}$; $k_+^2 k_z \mathbf{z}$ |

**Table Caption.** Basis functions for the various representations corresponding to hexagonal point group symmetry (after Yip and Garg [54]).

**Table V**
**Phenomenological Models for UPt$_3$**

1. 2D Representation
     (a) E$_{1g}$ (d-wave)
     (b) E$_{2u}$ (f-wave)

2. Nearly Degenerate Representations
     (a) 1D, 1D
     (b) 1D, 2D
     (c) d-wave (1D, 2D, 2D)

3. Spin Degeneracy
     (a) 3 component (**x**,**y**,**z**)
     (b) 2 component (**x**,**y**)

4. Center of Mass Degeneracy [ **Q**=(0.5,0,0) ]
     (a) odd frequency pairing
     (b) antiferromagnetic background

**Figure Captions**

Figure 1. Splitting of specific heat transition in $UPt_3$ (from Fisher et al. [11]).

Figure 2. Phase diagram in $UPt_3$ :H vs. T (from Adenwalla etal. [12]).

Figure 3. Phase diagram in $UPt_3$ :P vs. T (from Boukny et al. [13]).

Figure 4. Temperature dependence of specific heat in $(U,Th)Be_{13}$ (from Ott et al. [23]).

Figure 5. Enhanced μSR line width below $T_{c2}$ in $(U,Th)Be_{13}$ (from Heffner et al. (1989) [25]).

Figure 6. Phase diagram of $(U,Th)Be_{13}$ (from Heffner et al. (1990) [25]).

Figure 7. Temperature dependence of μSR amplitudes in $CeCu_2Si_2$ showing competition between paramagnetic $A_1$ (superconducting) and magnetic $A_2$ phases (from A. Amato [31]).

Figure 8. μSR relaxation rate in $UPd_2Al_3$ showing coexistence of superconducting and magnetic phases (from A. Amato [31]).

Figure 9. Temperature dependence of the magnetic intensity M in $UPt_3$ for two pressures, showing $M^2 \propto (T_N -T)$. The inset shows the variation of the splitting $\Delta T_c$ of the two superconducting transition temperatures vs. $M^2$. (from Hayden et al. [42]).

Figure 10. Modulus of the order parameter versus azimuthal angle for $E_{1g}$ and $E_{2u}$ cases.